\documentclass{aastex63}

\def\a{\"a}
\def\o{\"o}
\def\u{\"u}
\def\A{\"A}
\def\O{\"O}
\def\U{\"U}

\def\gapprox{\lower.4ex\hbox{$\;\buildrel >\over{\scriptstyle\sim}\;$}}
\def\lapprox{\lower.4ex\hbox{$\;\buildrel <\over{\scriptstyle\sim}\;$}}
\def\ref#1{\par\noindent\hangindent1cm {#1}}

\begin{document}

\title{Global Energetics of Solar Flares. 
XII. Energy Scaling Laws}

\correspondingauthor{Markus J. Aschwanden}
\email{aschwanden@lmsal.com}

\author{Markus J. Aschwanden}

\affiliation{Solar and Stellar Astrophysics Laboratory (LMSAL),
 Palo Alto, CA 94304, USA}

\begin{abstract}
In this study we test 18 versions of 5 fundamental
energy scaling laws that operate in large solar flares. 
We express scaling laws in terms of the magnetic potential field 
energy $E_p$, the mean potential field strength $B_p$,
the free energy $E_{free}$, the dissipated magnetic flare
energy $E_{diss}$, the magnetic length scale $L$, 
the thermal length scale $L_{th}$, the
mean helically twisted flux tube radius $R$, the sunspot 
radius $r$, the emission measure-weighted flare temperature $T_e$,
the electron density $n_e$, and the total emission measure $EM$,
measured from a data set of 173 GOES
M- and X-class flare events. The 5 categories of
physical scaling laws include  
(i) a scaling law of the potential-field energy, 
(ii) a scaling law for helical twisting, 
(iii) a scaling law for Petschek-type magnetic 
reconnection, (iv) the Rosner-Tucker-Vaiana scaling law, 
and (v) the Shibata-Yokoyama scaling law. We test the 
self-consistency of these theoretical scaling laws with 
observed parameters by requiring two criteria: a 
cross-corrleation coefficient of CCC$>$0.5 between the 
theoretically predicted scaling laws and observed values, and 
a linear regression fit with a slope of $\alpha \approx 1$
within one standard deviation $\sigma$. 
These two criteria enable us (i) to corroborate some 
existing (or modified) scaling laws, (ii) to reject other
scaling laws that are not consistent with the observations,
(iii) to probe the dimensionality of flare geometries,
and (iv) to predict various energy parameters based on
tested scaling laws.

\end{abstract}


\section{Introduction}

Scaling laws describe physical models with functional 
(mathematical) relationships between physical quantities 
that scale with each other over a significant parameter range,
often in form of power laws, e.g., 
$X = a^\alpha b^\beta c^\gamma$, where $X, a, b, c$ are
physical quantities, and $\alpha, \beta, \gamma$ are
power law exponents. The choice of relevant parameters 
($X, a, b, c$) requires the knowledge of specific
physical models, while the power law exponents 
($\alpha, \beta, \gamma$) can either be predicted by a 
physical model, or be obtained from forward-fitting 
of purely mathematical (power law) functions. Obviously,
since the latter method provides a physics-free 
parameterization only, it should be discouraged 
in favor of fitting specific physical scaling law models. 

A tricky player in any 
forward-fitting method is the data noise and the 
uncertainty of measured observables. 
The larger the data noise level is, the larger is the 
resulting uncertainty of the fitted power law exponents.
Vice versa, the best-fit power law exponents may be strongly biased
if some physical parameters are dominated by noise. 
In this study we use analytically derived scaling laws and
self-consistency tests with observations, rather than
fitting empirical power law exponents that by themselves 
do not reveal the underlying physical processes. 
The identification of relevant scaling laws often requires
3D models of observed 2D features (e.g., review of Aschwanden 2010).
One of the motivations of this study is the identification
of physical parameters that are most relevant for solar
flare prediction (machine-learning) methods. 

There are numerous studies on scaling laws of solar flares and
coronal heating, of which we mention here a few representative 
examples:
The Rosner-Tucker-Vaiana law applied to solar flares 
(Rosner et al.~1978; Aschwanden and Shimizu 2013);
scaling laws of electron time-of-flight distances and loop lengths 
(Aschwanden et al.~1996);
scaling laws of soft X-ray measurements in solar flares (Garcia 1998);
scaling laws of the coronal magnetic field and loop length 
in solar and stellar flares (Yokoyama and Shibata 1998; 
Shibata and Yokoyama 1999, 2002; Yamamoto et al.~2002; 
Aschwanden et al.~2008; Namekata et al.~2017); 
scaling laws for a nanoflare-heated solar corona 
(Vekstein and Katsukawa 2000); 
scaling laws from solar nanoflares to stellar giant flares
(Aschwanden 2007);
scaling laws and occurrence frequency distributions in self-organized 
criticality models of solar flares (Morales and Charbonneau 2008); 
universal scaling of space and temporal parameters in solar flares
(Aschwanden et al.~2013);
scaling laws of quasi-periodic pulsation periods and flare durations 
(Pugh et al.~2019);
scaling law between solar energetic particle events and solar flares 
(Kahler 2013).
 
In this study we are using a total of $\approx 20$ 
physical parameters, mostly unmeasured before,
with large statistics (173 flare events),
with detailed descriptions in a series of 11 papers on 
the global energetics of solar flares. For the first
time we apply energy scaling laws in terms of the
magnetic potential energy, the free energy, and the
dissipated magnetic energy in flares to rigorous
self-consistency tests. We are testing 
18 trial scaling laws of solar flares and find self-consistency
with observed data in at least one of each of the five
categories, but find that the best-fit cases depend
sensitively on the dimensionality of the geometric flare models.
The main goal of this study is to corroborate 
existing (or modified) scaling laws by means of quantitative criteria, 
as well as to reject other scaling laws that are not consistent 
with observations. Rigorous consistency tests of flare
scaling laws are largely absent in the literature. 
We refine pre-existing flare scaling
laws by varying the dimensionality of flare geometries
(such as coronal 3-D flare domains, 2-D chromospheric flare areas 
with a small vertical extent, and chromospheric 1-D flare ribbons. 
The scaling laws that pass our tests may reveal which
energy flare parameters are most relevant for
machine-learning data sets in flare-prediction efforts.

Data analysis and the results of the scaling laws are 
presented in Section 2, a Discussion of earlier work 
in the light of the new results in Section 3, and
Conclusions in Section 4. 

\section{ Observations and Methods }

\subsection{	Observations		}

The data analysis presented here is mostly based on
the measurements of flare parameters published in 
a series of 11 papers on the global energetics in solar
flares and associated {\sl coronal mass ejections (CMEs)},
using observations made with the {\sl Helioseismic
and Magnetic Imager (HMI)} (Scherrer et al.~2012) 
and the {\sl Atmospheric Imaging Assembly (AIA)}
(Lemen et al.~2012) onboard the {\sl Solar Dynamics
Observatory (SDO)} (Pesnell et al.~2012). In particular
we are using magnetic flare parameters of all
GOES M- and X-class flares during the first 3.5 years
of the SDO mission (2010-2014), which are obtained from 
173 flare events that occurred near the solar disk center 
(within a longitude range of $\pm 45^\circ$ due to 
magnetic modeling restrictions), measured in 
Paper I (Aschwanden, Xu, and Jing 2014a),
in Paper X (Aschwanden et al.~2020),
and in Paper XI (Aschwanden et al.~2019a).

Magnetic modeling with the {\sl Vertical Current Approximation
Non-Linear Force-Free Field (VCA-NLFFF)} code (Aschwanden 2013), 
which has the capacity of fitting coronal loops in the force-free 
coronal regime, rather than using the non-forcefree photospheric magnetic
vector field of other NLFFF codes (e.g., Wiegelmann et al.~2006), 
yields the following parameters (also called observables here):
the mean potential field $B_p$, 
the mean nonpotential field $B_{np}$,
the mean azimuthal magnetic field component $B_{free}$ 
associated with the free energy,
the mean dissipated magnetic field during a flare $B_{diss}$, 
the mean potential energy $E_p$, 
the mean nonpotential energy $E_{np}$,
the mean free energy $E_{free}$, 
the mean dissipated magnetic energy $E_{diss}$, 
the mean helical twist radius $R$,
the sunspot radial half width $r$,
the depth of buried magnetic charges $d$,
the mean magnetic length scale $L$,
and the flare duration $\tau_{flare}$. 
For testing the RTV scaling law we use five
additional parameters: The length scale of
thermal flare emission $L_{th}$, the
emission measure-weighted electron temperature $T_e$, 
the mean electron density of thermal flare emission $n_e$,
the total thermal emission measure $EM$, and the
total thermal energy $E_{th}$.
The minimum and maximum values, as well as the
means and standard deviations of these observables
of our data set of 173 analyzed events are 
given in Table 1. The observables will be used here to test
scaling law relationships in solar flares.
The main difference to previous work is the
exclusive testing of scaling laws that are based on
specific physical models (of solar flares), rather
than determining empirical scaling parameters.

\subsection{    Method of Testing Scaling Laws }

In this study we aim to identify the physical
parameters that are relevant for solar flare and coronal
heating processes, as well as to quantify which 
physical scaling laws are consistent with observations.
For self-consistency tests between theoretically predicted and
observationally measured scaling laws we use two criteria:
(i) a cross-correlation coefficent of CCC$>$0.5 between
theoretical and observational (logarithmic) parameter
values, and (ii) a power law slope of $\alpha \approx 1$ of the
linear regression fit between the (logarithmic) theoretical
$[log(Y)]$ and observational $[log(X)]$ parameters, within the
statistical uncertainties $\sigma$ of the linear regression fits,
\begin{equation}
        log[Y(X)] = const + (\alpha \pm \sigma) \ log[X] \ ,
\end{equation}
A standard linear regression fit is calculated, by
minimizing the chi-square error statistic of the dependent
variable $log[Y(X)]$ as a function of the independent 
variable $log[X]$ (Press et al.~1986). 
For instance, $X=E_p\propto B_p^2/8\pi$ 
is the potential magnetic field energy density 
calculated from a magnetogram ${\bf B}_p(x,y,z)$, while the peak
value $Y=max(B_p)$ represents a dependent variable, since 
it is derived from the same magnetogram. For the linear
regression fit we use the algorithm
LINFIT.PRO available in the {\sl Interactive Data Language
(IDL)} library. Successful cases fulfill the criteria 
CCC$>$0.5 and $|\alpha-1| \lapprox \sigma$ (see Table 2
and Figs.~4-6 and 8-9). We note that such self-consistency 
tests of scaling laws are generally lacking in literature.

\section{	Data Analysis and Results	} 

\subsection{ Scaling Law of the Potential-Field Energy }

The potential field represents the minimum energy state of a 
magnetic field (Priest 1975), and thus constitutes also a 
lower limit of the non-potential magnetic field energy.
The total magnetic energy $E_p$ of a potential field 
${\bf B}_p({\bf x})$ is generally expressed by an 
integral of the magnetic energy density $(B^2/8\pi)$ over a 
well-defined spatial volume $V$, 
\begin{equation}
        E_p = \int \left(
         { B_p^2(x,y,z) \over 8 \pi } \right) \ dV \quad
         = \int \left( { B_p^2(x,y,z) \over 8 \pi }
         \right) \ dx \ dy \ dz \ .
\end{equation}
This potential energy $E_p$ can be calculated from the magnetic 
field at the lower (photospheric) boundary $z_{phot}(x,y)$ 
of the computation box, such as given by a line-of-sight 
magnetogram $B_z(x,y,z=z_{phot})$, and from a potential-field 
extrapolation, which produces a volume-filling 
magnetic field ${\bf B}({\bf x}) = [B_x(x,y,z), B_y(x,y,z), 
B_z(x,y,z)]$.  

Potential field calculation methods include the Green's function
(Schmidt 1964; Sakurai 1982), the spherical Schmidt method, 
based on an eigen-function (spherical harmonic) expansion method,
using Legendre polynomials (Altschuler and Newkirk 1969), 
or simply by an iterative decomposition of (sub-photospheric)
unipolar magnetic charges (e.g., Aschwanden and Sandman 2010).
The latter approach is conceptually the simplest method, where
a potential field can be represented by a sum of unipolar
magnetic charges, whose field strengths fall off according to the
square-distance law (as it occurs also in electric or
gravitational potential fields),
\begin{equation}
	{\bf B}_p({\bf x}) = \sum_{j=1}^{N_m} {\bf B}_j({\bf x})=
	\sum_{j=1}^{N_m} B_j \left({r_j \over d_j}\right)^{-2}
	\left( {{\bf r}_j \over r_j} \right) \ ,
\end{equation}
where $N_m$ is the number of unipolar magnetic charges,  
${\bf r}_j = [(x-x_m), (y-y_m), (z-z_m)]$ is the vector of the
magnetic field direction, $[x_m, y_m, z_m]$ is the location
of the buried unipolar magnetic charge (in a Cartesian coordinate
system), $d_j$ is the depth
of the subphotospheric magnetic charge, and $B_j$ is the
magnetic field strength at the (photospheric) solar surface,
in vertical direction above the location of the unipolar
magnetic charge. The square-distance fall-off of a potential
field is equivalent to a divergence-free field, i.e.,
$\nabla \cdot {\bf B} = 0$.

We derive a simple scaling law for the potential energy $E_p$
of an active region in the solar corona, as a function of the
maximum magnetic potential-field strength $B_p$, the length scale
$L$ of the magnetic flare area, and the sunspot radius $r$. 
The potential field energy $E_p$ (Eq.~2) can be approximated by 
the mean magnetic field strength $B_p$ in a flare volume $V$,
\begin{equation}
	E_p = q_{norm} \ \left({ B_p^2 \over 8 \pi }\right) \ V \ , 
\end{equation}
with $q_{norm}$ being a normalization factor (between the
theoretically predicted and observed scaling law).
For the volume $V$,
which generally has a complex topology and geometry, we test
three basic trial geometries: a 3D cube geometry $V \propto L^3$,
a 2D (projected flare) area geometry $V \propto L^2 H$ with area 
$A=L^2$ and height extent $H$, and a 1D ribbon geometry 
$V \propto L H^2$ with length $L$ and cross-section
$H^2$ (Fig.~1). Thus we have 3 candidates for the scaling law of the
potential energy $E_p$, to which we refer as model EP-3D (cube), 
EP-2D (area), and EP-1D (ribbon), 
\begin{equation}
	E_p = q_{norm} \left( {B_p^2 \over 8\pi} \right)  
	\left\{
    	\begin{array}{ll}
	L^3             & \qquad {\rm Model\ EP-3D} \\
	L^2 H           & \qquad {\rm Model\ EP-2D} \\
	L^1 H^2         & \qquad {\rm Model\ EP-1D} \\ 
	\end{array}
	\right.  \ ,
\end{equation}
 
The height $H$ of the magnetic flare volume can be estimated 
from the location where the magnetic energy drops to half of the 
peak value, which is constrained by the distance-square law of 
the magnetic field $B(H)$ and the mean depth $d$ where 
the unipolar magnetic energies are buried (see Figs.~2 and 3),
\begin{equation}
	B_p(H) = B_0 \ \left( {d + H \over d} \right)^{-2}
	     = \left( {B_0 \over \sqrt{2}} \right) \ ,
\end{equation}
which yields the relationship, 
\begin{equation}
	H = (2^{1/4} - 1) \ d \approx 0.189 \ d \ .
\end{equation}  
Note that the magnetic field $B_p(H)$ drops over this height 
range $H$ from $B_0$ by a factor of $B_0/\sqrt{2}$, while the 
corresponding magnetic energy drops by a factor of $E_{p,0}/2$ (Fig.~3). 

Now we need to estimate the depth $d$ of the buried magnetic charges.
The depth $d$ is approximately equal to the sunspot radius $r$,
at least for the largest magnetic charge in an active region
(Aschwanden and Sandmann 2010), as we derive in the following. 
The radial magnetic field 
component is given by the square law, $B_r = B_0(d/\rho)^{-2}$,
with $\rho^2=r^2+d^2$ (see Fig.~2), which implies a line-of-sight 
component $B_z$ of
\begin{equation}
	B_z(r) = B_r(r) \cos{\theta} = B_p \left( { d \over \rho} \right)^3 
	= {B_p \over 2} \ ,	
\end{equation}
yielding a relationship between the depth $d$ and the sunspot
radius $r$ (Fig.~2), 
\begin{equation}
	d = {1 \over \sqrt{2^{2/3}-1}} \ r \approx 1.30 \ r \ .
\end{equation}

Substituting the magnetic height parameter $H$ with the
measured observable of the sunspot radius $r$ according
to the scaling of Eqs.~(7) and (9), i.e., $H \approx 0.25\ r$, 
we obtain the following (candidate) scaling laws for the 
potential field energy $E_p$ (in cgs-units),
\begin{equation}
	E_p \approx q_{norm} \ B_p^2 
	\left\{
    	\begin{array}{ll}
	0.0398\ L^3       & \qquad {\rm Model\ EP-3D} \\
	0.0098\ L^2\ r    & \qquad {\rm Model\ EP-2D} \\
	0.0024\ L\ r^2    & \qquad {\rm Model\ EP-1D} \\ 
	\end{array}
	\right.  \ ,
\end{equation}

We can now compare the observed potential energies $X=E_p^{obs}$, 
as calculated from the data of 173 M- and X-class flares in 
Aschwanden et al.~(2014a, Paper I) and Aschwanden et al.~(2019a; 
Paper XI), with the theoretically predicted scaling law candidates
$Y=E_p^{theo}$ (Eq.~10), using the observed values of the mean 
magnetic field strength $B_p$, magnetic length scale $L$, and 
sunspot radius $r$ of the 173 flare events. We show scatter plots 
$Y(X)$ of these two quantities $X$ and $Y$ in Fig.~(4), where we find 
that the best case is the 2D model (Fig.~4b), with a cross-correlation 
coefficient of CCC=0.69, a slope of
$\alpha=1.11\pm0.09$ for the linear regression fit of the logarithmic values,
i.e., $log(Y)=const + (\alpha\pm\sigma) \times log(X)$, and a normalization 
factor $q_{norm}=6.3$. 
The best-fit model (2D) represents the 2D (projected flare) area model,
\begin{equation}
	E_p \approx 0.0617\ B_p^2 \ L^2 r \quad {\rm (erg)} \ .
\end{equation}
Note that the flare area $A=L^2$ was measureded from magnetic parameters
in an earlier study (Aschwanden et al.~2014a; Paper 1),
by accumulating the flare area above a threshold $B_{free}
=B_{\varphi} \ge 100$ G of the dissipated flare energy $E_{diss}(x,y,t)$,
yielding a cumulative flare area that monotonically grows during the
flare duration.

In order to assess the consistency between theoretical 
scaling laws and observed values we use two different 
critieria (Section 2.2):
The cross-correlation coefficient CCC, and the
linear regression slope $\alpha\pm\sigma$, which in the 
case of a perfectly matching model should converge
to unity, i.e., CCC=1.0 and $\alpha=1.0$. 
Investigating the three trial models (Eq.~10), we find 
that both the 3D cube geometry (3D model) and the 
1D ribbon geometry model are not consistent with
the observed data, while the 2D flare area model produces
a slope $\alpha=1.11\pm0.09$ that is consistent with the
data within about one standard deviation. For the normalization
factor we find a value of $q_{norm} \approx 6.3$,
which can be interpreted in terms of the equivalent number of 
large unipolar magnetic charges. Small flares exhibit generally at least
a dipolar magnetic configuration ($N_{magn}=2$), while large
flares are generally magnetically more complex and have a
quadrupolar configuration ($N_{magn}=4$), so that we can easily
explain more complex structures with $N_{magn} = q_{norm} = 6.3$
for the sample of largest flares (GOES M- and X-class events)
used in this sample here.

\subsection{ Scaling Law of Helical Twisting }

Many physical models of flares employ untwisting of stressed
magnetic fields as a primary trigger of a flare event. 
In the simplest version, a potential field $B_p$ becomes
helically twisted in the preflare phase, up to a maximum
twist that can be quantified with a non-potential field
$B_{np}$, before it becomes unstable, e.g. due to the
kink instability or torus instability, and consequently 
releases or dissipates a part $E_{diss}$ of the free energy $E_{free}$, 
\begin{equation}
	E_{diss} \lapprox E_{free} = E_{np} - E_p \ .
\end{equation}
Because of the quadratic dependence of the magnetic
energy on the magnetic field strength, i.e., 
$E \propto B^2/8\pi$, we have a Pythagorean relationship
between the three magnetic field components
$[B_{np}, B_p, B_{free}]$, namely
\begin{equation}
	B_{free}^2 = B_{np}^2 - B_p^2 \ .
\end{equation}
which also provides a vector equation for the directions
of the magnetic field components,
\begin{equation}
	{\bf B}_{free} = {\bf B}_{np} - {\bf B}_p \ .
\end{equation}
For the case of a helically twisted field, which can most
conveniently be expressed with spherical coordinates
$(r, \varphi, \theta)$, the potential field component
is aligned with the radial field $B_r$, and the field component
associated with the free energy contains the azimuthal
twist $B_{\varphi}$, which is perpendicular to the
radial component (by applying the Pythagorean relationship).
The helical twist involves a partial rotation of a cylindrical
flux tube around its symmetry axis, which can be expressed
by the twist angle $\mu$ relative to the untwisted potential
field,
\begin{equation}
	\tan{\mu} = \left( {B_{\varphi} \over B_p} \right)
	= \left({ E_{free} \over E_p} \right)^{1/2} 
	= \left( 2 \pi N_{twist} {R \over L} \right) \ ,
\end{equation}
where $N_{twist}$ is the number of full twisting turns,
$R$ is the mean radius of a helically twisted flux tube, 
and $L$ is the mean loop (or flux tube) length.
Thus, we can formulate a scaling law for the free energy
$E_{free}$ as a function of the observables 
[$E_p, N_{twist}, L, R$], 
\begin{equation}
	E_{free} = E_p
	  \left( 2 \pi N_{twist} {R \over L} \right )^2 \ .
\end{equation}
Inserting the scaling of the potential energy $E_p$ (Eq.~10)
into this expression (Eq.~16) we obtain a scaling law as
a function of the observables $[B_p, L, R, r]$. 
We test three geometric models
of scaling laws by comparing with the observed free energy
$E_{free}$ (Fig.~5), 
\begin{equation}
        E_{free} = E_p \left( 2 \pi N_{twist} {R \over L}\right)^2 
	= q_{norm} B_p^2\ R^2 \
	\left\{
        \begin{array}{lll}
        0.0353\   L             &\qquad {\rm Model\ EF-3D} \\
        0.00873\  r             &\qquad {\rm Model\ EF-2D} \\
        0.00216\  L^{-1}\ r^2   &\qquad {\rm Model\ EF-1D} \\
	\end{array} 
	\right.  \ .
\end{equation}

The number of full turn twists has 
been found to have a mean value of $N_{twist} \approx 0.14\pm0.03$ 
in an earlier study (Aschwanden 2020). 
Fig.~5 shows that the best-fit model occurs for the 1D flare 
ribbon geometry.
The theoretical 1D model shown in Fig.~5c has a high 
cross-correlation coefficient (CCC=0.86), produces a 
slope of $\alpha=1.04\pm0.05$ for the linear regression fit,
a normalization factor of $q_{norm} = 165$. The best-fit case
can be written as,
\begin{equation}
	E_{free} = 0.356 \ B_p^2\ R^2\ r^2\ L^{-1} 
	\qquad {\rm (erg)} \ .
\end{equation} 
The normalization factor of $q_{norm} \gg 1$
can be interpreted as the number of cross-sectional loop
strands that are constrained by the observed total
free energies.

The benefit of this result (Eq.~18) is that the free energy, 
which is generally close to the magnetically dissipated flare
energy, $E_{diss} \lapprox E_{free}$, can simply be estimated from 
the peak value of the potential field $B_p$, the loop length
scale $L$, the twist radius $R$, and the sunspot radius $r$. 

\subsection{ Scaling Law for Petschek-Type Magnetic Reconnection }

A common configuration of magnetic reconnection processes in
the lower solar corona involves the interaction between two (more or
less) vertical field lines with opposite magnetic polarity, which
exchange their connectivity. The geometry is characterized by an
X-point, where horizontal plasma inflows with low speed is
transported through the diffusion region and produces accelerated
Alfv\'enic outflows in downward and upward direction (Fig.~7).
We can characterize the diffusion region with a typical length
scale $L$ in horizontal, vertical, and line-of-sight direction.
Assuming that sideward inflows last for the duration $\tau_{flare}$
of a flare, starting at a distance of $L/2$ from the center of the
X-point (Fig.~7, left panel), we can estimate the inflow velocity as,
\begin{equation}
	v_1 = {L/2 \over \tau_{flare}} \ , 
\end{equation}
which has mean value of $v_1=22$ km s$^{-1}$, according to our
measurements of $L=40\pm25$ Mm for the magnetic length scale and
$\tau_{flare}=900$ s for the flare duration (Table 1).
The magnetic energy $E_{diss}$ dissipated during a flare can be 
expressed by the magnetic energy that is processed through the
diffusion region with a volume of $V=L^3$ and an inflow rate
of $dV/dt=L^2 v_1$, which yields with Eq.~(20), with consideration
of doubling the half flare volume on either side of the X-point,
\begin{equation}
	E_{diss} = \int \left( {dE \over dt} \right) dt
	= 2 \left( {B_{free}^2 \over 8\pi }\right) {dV \over dt} \tau_{flare}
	=   \left( {B_{free}^2 \over 4\pi}\right) L^2 v_1 \tau_{flare}
	=   \left( {B_{free}^2 \over 8\pi }\right) L^3 \ ,
\end{equation}
where $B_{free}$ is the magnetic field strength associated with the
free energy. From the definition of the free energy $E_{free}$ in terms of
the difference between the nonpotential $E_{np}$ and the potential energy
$E_p$ (Eq.~12), we can determine the mean value of $B_{free}$ from the
(observed and measured) energy ratio,  
\begin{equation}
	B_{free} = B_p \ \left( { E_{free} \over E_p } \right)^{1/2} \ .
\end{equation}
Therefore, the magnetic reconnection scenario of Petschek predicts
a scaling law of $E_{diss} \propto B_{free}^2 L^3$ (Eq.~20) that is based
on the observables [$B_{free}, L$]. 

We explore three different geometries as variants of the tested
scaling laws: 
\begin{equation}
        E_{diss} = q_{norm} \ {B_{free}^2 \over 8\pi}\ \left\{
        \begin{array}{ll}
        L^3                         & \qquad {\rm Model\ ED-3D} \\
        L^2\ H = 0.00983\ L^2\ r    & \qquad {\rm Model\ ED-2D} \\
        L\ H^2 = 0.00243\ L\ r^2    & \qquad {\rm Model\ ED-1D} \\
	\end{array} 
	\right.  \ .
\end{equation}
These three models include cubic volume flare geometries 
$E_{diss} \propto L^3$ (Figs.~6a); 
projected flare area geometries
$E_{diss} \propto L^2$ (Figs.~6b);
and (ribbon-like) linear flare geometries  
$E_{diss} \propto L$ (Figs.~6c).

We show this scaling law in Fig.~6, compared with the observed values
of $E_{diss}$. The best-fit case (model 3c) exhibitis a cross-correlation 
coefficient of CCC=0.84 for the dissipated energy $E_{diss}$,
a slope of the linear regression fit $\alpha=1.08\pm0.05$, and a
normalization factor of $q_{norm}=242$. Thus the best-fit scaling law
indicates the 1D flare ribbon geometry of the volume of dissipated magnetic
energy, rather than a 2D flare area or a 3D cube geometry.
This best-fit model can be expressed as,   
\begin{equation}
        E_{diss} \approx 0.588\ B_{free}^2\ L\ r^2  
                \quad ({\rm erg}) \ .
\end{equation}
The normalization factor $q_{norm}=242$ indicates a multi-strand
configuration, similar as we found for the best-fit model of
helically twisted loops (model 2c).  
The implementation of the Petschek model
in the tested scaling laws implies that the horizontal
inflow of plasma into the reconnection region with speed
$v_1 = (L/2) \tau_{flare}$ and volume $V \propto L\ r^2$
produce a scaling law that is consistent with the observed 
data (Fig.~6c).

\subsection{ The Rosner-Tucker-Vaiana Scaling Law }

The Rosner-Tucker-Vaiana (RTV) scaling law (Rosner et al.~1978)
has originally been designed to quantify a steady-state
solution of a coronal loop, where the volumetric heating rate
is balanced by the conductive and radiative loss rate.
In addition, the RTV scaling law has been successfully
applied to solar flares also, which consist of loop assemblies 
with complex magnetic topologies and geometries. Although the 
dynamics of flare loop systems is certainly not quasi-stationary 
during a flare, there is a critical equilibrium point
at the flare peak time where heating
and (conductive and radiative) cooling are approximately
in balance, while the pre-flare phase is dominated by heating,
and the post-flare phase is dominated by cooling.  

The analytical solution can be expressed by four physical
variables: the loop length of the thermal plasma $L_{th}$, 
the maximum temperature 
$T_{max}$, the pressure $p_0=2 n_e k_B T_{max}$, and the
volumetric heating rate $E_{H0}$ (Rosner et al.~1978), 
\begin{equation}
	T_{max} \approx 1400 (p_0\ L_{th})^{1/3} \ ,
\end{equation}
\begin{equation}
	E_{H0} \approx 0.98 \times 10^5
	p_0^{7/6} L_{th}^{-5/6} = 0.95 \times 10^{-6} T_{max}^{7/2} L_{th}^{-2} \ .
\end{equation}
An equivalent formulation of the RTV law that contains the electron
density $n_e$, emission measure $EM$, and multi-thermal energy $E_{th}$
has been given in Aschwanden and Shimizu (2013) and in 
Aschwanden et al.~2015 (Paper II), 
\begin{equation}
	T_{RTV} = 1.1 \times 10^{-3} \ n_e^{1/2} L_{th}^{1/2} \ ,
\end{equation}
\begin{equation}
	n_{RTV} = 8.4 \times 10^{5} \ T_e^{2} L_{th}^{-1} \ ,
\end{equation}
\begin{equation}
	L_{RTV} = 8.4 \times 10^{5} \ T_e^{2} n_e^{-1} \ ,
\end{equation}
\begin{equation}
	EM_{RTV} = 1.5 \times 10^{12} \ T_e^{4} L_{th} \ ,
\end{equation}
\begin{equation}
	E_{th,RTV} = 7.3 \times 10^{-10} \ T_e^{3} L_{th}^2 \ .
\end{equation}
We are testing now these five scaling law relationships
separately in Fig.~8. Most of them have a high
cross-correlation coeffient in the range of CCC=0.34-0.90,
which corroborates the validity of the RTV scaling law
The best match is found for the RTV scaling law of the
thermal energy $E_{th,RTV} \propto T_e^3 L_{th}^2$, with
CCC=0.90 and $\alpha=0.95\pm0.04$ (Fig.~8e).

The five scaling parameters are five different
formulations of the same RTV scaling law and thus are
redundant, but they demonstrate that the accuracy in
the determination of a scaling law improves with
larger parameter ranges: the thermal energy has
the largest range of about three orders of magnitude
and exhibits the highest correlation coefficient 
(CCC=0.90) and a slope $\alpha=0.95\pm0.04$ near unity (Fig.~8e), 
while the temperature has a range of less than one order
of magnitude, with CCC=0.34 and $\alpha=0.95\pm0.04$ (Fig.~8e).
The uncertainty of measurements that restrict the
accuracy of detected scaling laws is expected to
improve with larger parameter (in particular temperature) 
ranges (e.g., by including GOES C-class events, besides the M- and 
X-class flares sampled here).

\subsection{ Shibata-Yokoyama Scaling Laws	}

Assuming that the energy release in solar or stellar
flares is dominated by magnetic reconnection and
conductive energy loss, ``universal scaling laws'' 
were derived that roughly agree with MHD simulations
(Yokoyama and Shibata 1998; Shibata and Yokoyama 
1999, 2002). The analytical derivation of these
scaling laws is based on four assumptions: (i)
The flare temperature at the loop apex is balanced
by the conduction cooling rate and the volumetric
heating rate $(Q=d/ds(\kappa_0 T^{5/2} dT/ds) \approx 
(2/7) \kappa_0 T_e^{7/2}/L_{th}^2)$, with the Spitzer
conductivity constant $\kappa_0$, yielding a flare
loop apex temperature of,
\begin{equation}
	T_e \approx \left({2 Q L_{th}^2 \over \kappa_0}
	\right)^{2/7} \ ,
\end{equation}
(ii) the heating rate $Q$ is given by the reconnection
rate in Petschek's theory (with inflow velocity $v_{in}$,
outflow (Alfv\'enic) velocity $v_A$, and shock
inclination angle $\sin{\theta}=v_{in}/v_A$
equivalent to the Alfv\'enic Mach number $M_A=v_{in}/v_A$,
\begin{equation}
	Q = \left( {B^2 \over 4\pi} \right)  
	\left({v_{in} \over v_A}\right)
	\left({1 \over \sin{\theta}} \right)
	\approx \left( {B^2 \over 4\pi} \right)  
	\left({v_A \over L_{th}}\right) \ ,
\end{equation}
(iii) the upper limit of the gas pressure is given by
the magnetic energy density,
\begin{equation}
	p_{th} = 2 n_e k_B T_e = p_m = \left( {B^2 \over 8\pi}
	\right) \ ,
\end{equation}
and (iv) the volume of the heated flare plasma 
is approximated by a cubic geometry with total
emission measure $EM$ of, 
\begin{equation}	
	EM \approx n_e^2 L_{th}^3 \ .
\end{equation}
Using these four assumptions, four scaling laws can be
expressed explicitly for the parameters $T_e$, $EM$,
$B$, $L_{th}$ and $n_e$ (Shibata and Yokoyama 1999, 2002),
\begin{equation}
	T_e = 3 \times 10^7 \ 
	\left({ B \over 50\ {\rm G}}\right)^{6/7}  
	\left({ n_e \over 10^9\ {\rm cm}^{-3}}\right)^{-1/7}  
	\left({ L_{th} \over 10^9\ {\rm cm}}\right)^{2/7} \quad [{\rm K}] \ ,
\end{equation} 
\begin{equation}
	EM = 10^{48} \  
	\left({L_{th} \over 10^9\ {\rm cm}}\right)^{5/3}  
	\left({n_e \over 10^9\ {\rm cm}^{-3} }\right)^{2/3} 
	\left({T_e \over 10^7\ {\rm K} }\right)^{8/3} \quad [{\rm cm}^{-3}] \ ,
\end{equation} 
\begin{equation}
	B = 50 \ 
	\left({EM \over 10^{48 \ {\rm cm}^{-3}}}\right)^{-1/5} 
	\left({n_e \over 10^9\ {\rm cm}^{-3} }\right)^{3/10} 
	\left({T_e \over 10^7\ {\rm K} }\right)^{17/10} \quad [{\rm G}] \ ,
\end{equation} 
\begin{equation}
	L_{th} = 10^9 \ 
	\left({EM \over 10^{48 \ {\rm cm}^{-3}}}\right)^{3/5} 
	\left({n_e \over 10^9\ {\rm cm}^{-3} }\right)^{-2/5} 
	\left({T_e \over 10^7\ {\rm K} }\right)^{-8/5} \quad [{\rm cm}] \ .
\end{equation} 

These equations derived in Shibata and Yokoyama (1999), have been 
referred to as the {\sl pressure balance scaling law}, where the parameters
$[T_e, EM, B, L_{th}]$ are explicitly expressed as a function of these
physical parameters. 
We are testing now these four variants of the Shibata-Yokoyama scaling
laws in Fig.~9, using our SDO-based measurements of
$T_e$, $B \approx B_{np}$, $n_e$,
$L_{th}$, and $EM$. The best-fitting models are 
$EM \propto L^{5/3} n_e^{2/3} T^{8/3}$ 
with a cross-correlation coefficient of CCC=0.76 and a
linear regression slope of $\alpha=0.79\pm0.05$ (Fig.~9b), and 
$L_{th} \propto EM^{3/5} n_e^{-2/5} T^{-8/5}$ 
with a cross-correlation coefficient of CCC=0.87 and a
linear regression slope of $\alpha=1.18\pm0.05$ (Fig.~9d).  
The other variants indicate that the relative range
of parameter variations is extremely narrow for the scaling laws
of the temperature $T_e$ (Fig.~9a) and the magnetic field $B_{np}$
(Fig.~9c), which implies a large uncertainty in the linear
regression slope of these two parameters. We expect that this
selection bias can be overcome by extending the tested 
data sets to smaller (C-class) flares.  

\section{	Discussion }

\subsection{	Coronal Heating Scaling Laws	}

Since heating of coronal plasma occurs both in non-flaring
Quiet Sun regions as well as in flaring active regions, we 
comment on both processes in our discussion.

The most extensive compilation of scaling laws related to
coronal heating has been presented in the pioneering study of
Mandrini et al.~(2000), entailing a total of 22 models,
grouped into stressing (DC) and wave (AC) models.
These power laws have the form of 
$E_{heat} \propto B^\beta L^\lambda v^\gamma ...$, where 
$E_{heat}$ is the heating rate, $B$ is the mean magnetic field,
$L$ is a length scale, and $v$ is a transverse velocity
at the footpoints of coronal (magnetic) loops. The authors
find that the mean magnetic field $B$ scales with the length
$L$ of a magnetic field line by the relationship 
$log(B)=c_1+c_2 \log{L} + c_3/2 log(L^2+S^2)$, which can be
approximated by a double power law, with different slopes for 
the near-field and far-field (magnetic) regime, but 
cannot be represented by a single power law relationship
$B \propto L^\delta$. Nevertheless, reducing the data set
to loops with intermediate lengths and combining it with a
heating rate of $E_{heat} \propto L^{-2}$ (Klimchuk and
Porter 1995), a universal scaling law of 
$B \propto L^{-0.88\pm0.3} \propto L^{-1}$ has been postulated,
from which it was concluded that DC models are more consistent
with observations than AC models (Mandrini et al.~2000). 
In contrast to this result of $B \propto L^{-1}$, one would
expect that a potential field drops off with the square-distance 
law $B \propto L^{-2}$ (Eq.~3), at least for field lines that
originate at the strongest magnetic field location of an
active region, which is generally dominated by the leading sunspot. 
Furthermore, a scaling law of the heating rate should primarily
include the regions with the strongest magnetic fields, 
rather than magnetic field lines with (ill-defined) intermediate 
lengths, as employed in the study of Mandrini et al.~(2000). 

A novel approach to identify coronal heating mechanisms has been
presented by Schrijver et al.~(2004), by simulating the full-Sun
corona with an ensemble of 50,000 hydrostatic loops, subject to
a constant heating rate with the parameterization 
$F_{heat} \propto B^\beta L^\lambda$, for which a best-fit 
scaling law of 	
$F_{heat} \propto B^{1.0\pm0.3} L^{-1.0\pm0.5}$ was found.
Since this formulation using the Poynting flux $F_{heat}$
(in units of erg cm$^{-2}$ s$^{-1}$) relates to the
volumetric heating rate $\varepsilon_{heat} \propto F_{heat}/L$
(in units of erg cm$^{-3}$ s$^{-1}$), we would expect a
scaling law of $\varepsilon_{heat} \propto B^1 L^{-2}$
for the volumetric heating rate, and a scaling law of 
$E_{heat} \propto \varepsilon_{heat} L^3 \tau_{dur}
\propto B^{1} L^{1} \tau_{dur}$ (in units of erg) 
for the total dissipated energy. This hypothetical
scaling law is in marked contrast to other heating models
or flare models, which all have the basic relationship of
magnetic energies in terms of $E_{magn} \propto B^2 L^3$
(Eqs.~4, 11, 18, 23) in physical units of (erg) = (G$^2$ cm$^3$).
Schrijver et al.~(2004) point out that their result is
consistent with the allowed models in the study of 
Mandrini et al.~(2000), but the dependence of the heating
flux density $F_{heat}$ with $B/L$ is also consistent with
the study by D\'emoulin et al.~(2003), where a 
volumetric heating rate of $\varepsilon_{heat} \propto B^2$
is expected on grounds of the physical units anyway. 
Note that all magnetic scaling
models used here are based on the scaling of 
$E_{magn} \propto B^2$. Regarding self-consistency tests,
the simulations of Schrijver et al.~(2004) do show the
correlation between the observed soft X-ray fluxes and the
simulated fluxes $F_{heat}$, but do not demonstrate
whether the scaling law $F_{heat} \propto B/L$
is consistent with the data.

\subsection{	Helical Twist Scaling Laws	 	}

Free (magnetic) energy can be stored by helical twisting
of magnetic flux tubes. This twist can be measured in terms
of the number of full turns, where the free energy monotonically
increases with the twist angle (Eq.~16), up to a maximum
rotation angle that triggers the kink instability, which
occurs after about one full turn, i.e., $N_{twist}
\approx 1.0-1.5$ (Hood and Priest 1979, 1981; T\"or\"ok
and Kliem 2003; Kliem et al.~2010). From our derivation
of a twist-related scaling law (Section 2.2) we found
that the free energy depends mostly on the twisted
flux tube radius $R$ (Eq.~16), the magnetic field
strength $B_p$, the sunspot radius $r$, and the loop 
length $L$. The flux tube radius $R$ of helical
twist corresponds to the distance of detected twisted
field lines from the symmetry axis of the underlying
potential field, which amounts here to $R=12\pm12$ Mm
(Table 1), which is generally outside of the sunspot
radius of $r \approx 4.5\pm1.1$ Mm. 

How is the helical twist measured in the solar corona?
Fortunately, the VCA-NLFFF code parameterizes the magnetic
field into three orthogonal components, where the azimuthal
component $B_{free}=B_\varphi$ contains all information 
about the helical twist relative to the (untwisted) potential 
field component. This way we obtained in a previous
statistical study (Aschwanden 2019c) 
a twist number (or Gaussian linkage number)
of $N_{twist}=0.14 \pm 0.03$, with an absolute upper limit of
$N_{twist} \lapprox 0.5$, which is about a factor of two below
the kink instability. This somewhat lower (than expected) value may
indicate that no force-free field solution exists in that
regime of $N_{twist} \approx 0.5-1.0$. 

Agreement has been found between magnetic helicity estimations
and a twist number method, as calculated with the Berger-Prior
formula, which is supposed to be suitable for arbitrary geometry 
and both force-free and non-force-free models (Guo et al.~2017).
The twist number method has been applied to MHD simulations from a NLFFF code, 
over a range of $N_{twist} = 0.02-0.61$ (Guo et al.~2017) that is
similar to the results from our VCA-NLFFF code, i.e.,
$N_{twist} \lapprox 0.5$ (Aschwanden 2019c). 

\subsection{	Magnetic Reconnection Scaling Laws 	}

Soft X-ray and EUV emission of solar flares, observed during
the Yohkoh era, provided strong evidence for the existence
of magnetic reconnection processes. The theoretical modeling
focused first on the Sweet-Parker current sheet model
(Sweet 1958; Parker 1963, 1988), which was found to be
too slow to account for the impulsive flare phase, while a
more realistic concept was initiated with the Petschek model 
(Petschek 1964; Aschwanden 2020). A salient feature of the Petschek model
is the smaller size of the X-type current sheets, which
allows for a slow horizontal inflow (with speed $v_{in}
\approx 20$ km s$^{-1}$) and fast vertical Alfv\'enic 
outflows (with a typical speed of $v_A \approx 2000$ km s$^{-1}$),
which corresponds to an Alfv\'enic Mach number 
of $M_A=v_{in}/v_A\approx 0.01$, consistent with the expected
high Lundquist number of $R_m \approx 10^8-10^{12}$ in the solar corona. 
The related magnetic reconnection
rate $dV/dt \propto L^2 v_{in}$ with $v_{in}=(L/2)/\tau_{flare}$
constrains the volume $V$ of the dissipated magnetic energy in a flare, 
i.e., $E_{diss} \propto B_{free}^2 V$ (Eq.~23).
A multi-loop geometry of a flaring reconnection region can be quantified 
by the number ($q_{norm}$), the length ($L$), and the cross-sectional radius $(r)$
of a single loop strand. 
Our best-fit procedure does indeed match the observations 
with the 1D geometry $E_{diss} \propto q_{norm} B_{free}^2 L r^2$ 
(Fig.~6c).
 
Originally, Yokoyama and Shibata (1998) tested a Petschek-type
scaling law with MHD simulations, assuming that the scaling law
obeys the relationship $T_e \propto B^{6/7}$ (Eq.~35). 
A more complete derivation of Petschek-type flare scaling laws
was presented in Shibata and Yokoyama (1999), based on the
assumptions that flare heating is balanced by the conductive
cooling rate, and that the gas pressure is limited by the
magnetic energy density (Eq.~31-33). Quantitative evidence
of the Yokoyama-Shibata ``universal scaling law'' has been
demonstrated by the fact that Yohkoh-observed solar flares
are ``sandwiched'' in the emission measure-temperature 
$EM-T$ diagram inbetween solar microflares and stellar giant
flares (Fig.~1 in Shibata and Yokoyama 1999). However,
this matching of the the $EM-T$ diagram does not fully
demonstrate a cross-correlation between the observed
and theoretically predicted scaling laws. In this
study we tested the Yokoyama-Shibata scaling law with 173 flare
events observed with SDO and find that a good match
for two of the four Shibata-Yokoyama relationships, 
i.e., $L_{th} \propto EM^{3/5} n_e^{-2/5} T^{-8/5}$
(Fig.~9d), and $EM \propto L^{5/3} n_e^{2/3} T^{8/3}$
(Fig.~9b), while the two other scaling laws do not 
corroborate the expected scaling laws, most likely 
because the temperature $T$ and magnetic field
strength $B$ exhibit a very narrow parameter range
(see also standard devations in Table 1). Hence,
the parameter ranges have to be expanded by
including smaller flares (of GOES C-class and below)
to corroborate the validity of the Shibata-Yokoyama
scaling law. 

Shibata and Yokoyama (2002) generalized the observationally
established emission measure-temperature relationship 
$EM \propto T^4$ by combining different hydrostatic models,
such as the thermal conduction-driven model (Eq.~31), 
the magnetic-thermal pressure balance model (Eq.~33), or the
enthalpy-conduction balance model $(5 n_e k_B T C_s
\approx \kappa_0 T^{7/2}/L)$. Moreover, certain forbidden
regions in the EM-T diagram have been identified, for
instance when chromospheric evaporation sets in and the
thermal pressure overcomes the magnetic pressure so that
plasma confinement is not possible anymore, which violates
the assumption of the gas pressure being an upper limit
of the magnetic energy density (Eq.~33). We find that
our self-consistency criteria (CCC$\lapprox$1.0 and
$\alpha \approx 1$) provide powerful tests to sort out
which model is most consistent with the observed flare data. 

\subsection{	Thermal Flare Emission Scaling Laws	} 

We tested and discussed the RTV scaling law previously in 
Paper II (see Fig.~8 therein). 
The Rosner-Tucker-Vaiana (RTV) law represents a generalization
of the Shibata-Yokohama model on one side, based on the 
assumption that the heating rate $E_{heat}$ is balanced by 
both the conductive $E_{cond}$ and the radiative loss
rate $E_{rad}$, i.e., $E_{heat} = E_{cond} + E_{rad}$, 
while it is a reduction on the other side, by including
thermal parameters only and ignoring magnetic parameters.
Another special condition in the application of the RTV law 
is that the energy balance applies only to the peak time
of the flare, where the heating rate dominates over the
loss rates before the flare peak time (including the 
preflare phase), while conductive and radiative losses 
dominate the heating rate after the flare peak (fading
into the post-flare phase). We find that our self-consistency
tests with the flare data yield the best agreement for the
thermal energy parameter, $E_{th} \propto T_e^3 L_{th}^2$,
with CCC=0.90 and $\alpha=0.95\pm0.04$ (Fig.~8e). Other
RTV parameters, $[L_{th}, n_e, EM]$, show a significant
correlation (CCC$\gapprox$0.5), but exhibit a deviation from
the linear regression slope of $\alpha=1$, most likely 
because of the narrow temperature range, since the
temperature occurs in each of the five RTV scaling law
formulations show in Fig.~8 (Eqs.~26-30). In contrast,
the thermal energy displays the largest parameter range
of about 3 orders of magnitude 
($E_{th}\approx 10^{29}-10^{32}$ erg,
Fig.~8e). We learn from our analysis that it is of utmost
importance to cover a large parameter range in the
linear regression fits when testing scaling laws.
Larger parameter ranges that include C-, M-, and X-class
flares have been measured by Warmuth and Mann (2016a, 2016b),
for comparisons of thermal and non-thermal flare parameters.
However, since the occurrence frequency distributions of
solar flare parameters typically follow a power law 
distribution $N(x) \propto x^{-p}$ with a power law index of 
$p \approx 1.8$ (Crosby et al.~1991), completely sampled
parameter data sets become prohibitively large below 
the M- or C-class level. Fortunately, the testing of
scaling laws does not require complete sampling in
the calculation of cross-correlation coefficients. 

\section{	Conclusions		}

In this study we tested 18 variants of 5 basic physical
scaling laws that describe different aspects of solar flares,
which includes: (i) a scaling law of the potential-field
energy, (ii) a scaling law for helical twisting,
(iii) a scaling law for Petschek-type magnetic reconnection,
(iv) the Rosner-Tucker-Vaiana scaling law, and (v) the
Shibata-Yokoyama scaling law. We test the self-consistency
of these theoretical scaling laws with observed parameters
by requiring two conditions, a cross-corrlation coefficient
of CCC$>0.5$ between the observed and theoretically predicted
scaling laws, and a near-identical linear regression slope
$\alpha \approx 1$ for the observed values and theoretical scaling
laws. These two criteria are necessary but not sufficient
conditions to verify the existence and validity of
theoretical scaling laws. They are rarely quantified in the
literature. In the following we summarize the best-fit
solutions and their underlying physical assumptions,
using a data set of 173 GOES M- and X-class flare events.

\begin{enumerate}
\item{The potential field energy $E_p$ is a lower limit to 
any non-potential field, and is a near-constant (or at least 
a slowly-varying) parameter during a solar flare, which
makes it easier to determine than by using time-varying flare
parameters. Our assumptions imply that this 
potential field energy can be estimated from the peak 
of the magnetic field $B_p(x,y,z)$ in an active region with volume
$V=A H$, within a height range $H \approx 0.2\ d$ over which the 
potential field energy decreases by half, which relates to 
the depth $d$ of buried magnetic charges and the sunspot 
radius $r$ by $d \approx 1.3\ r$, yielding the scaling law
$E_p \propto B_p^2 L^2 r$ (Eq.~11). The 2D volume model of the
projected flare area $A=L^2$ appears to provide the best-fitting
geometric model for estimating the potential magnetic energy $E_p$,
according to CCC=0.69 and $\alpha=1.11\pm0.09$ (Fig.~4b). 
Thus we learn that the (volume-integrated) total magnetic
potential energy of a flaring active 
region needs to be approximated with a vertical extent that 
corresponds to the magnetic scale height $H$, rather than to use 
a cubic volume approximation.}

\item{An upper limit of the dissipated magnetic energy in a flare 
is the free energy, $E_{diss} \lapprox E_{free} = E_{np} - E_p$.
If the dissipated magnetic energy is entirely due to untwisting
of the helically twisted flare loops, a scaling law can be 
formulated as a function of the potential field energy $E_p$,
the number of full turn twist $N_{twist}$, the radius $R$, and the
length $L$ of helically twisted cyclinders, i.e., 
$E_{free} \propto E_p N_{twist}^2 (R/L)^2$ (Eq.~16), or 
$E_{free} \propto B_p^2 R^2 L^{-1} r^2$ (Eq.~18).
The best-fitting solution scaling law is found to be
most consistent with a 1D flare-ribbon geometry, 
according to CCC=0.86 and $\alpha=1.04 \pm 0.05$ (Fig.~5c), 
rather than with
a 2D area, or a 3D cube geometry. Hence we learn that the
free energy that possibly can be dissipated during a flare is
confined to the 1-D flare ribbons, rather than being uniformly
distributed in a coronal 3-D cube that encompasses a flaring
active region.}

\item{The Petschek-type magnetic reconnection in the
X-point of a magnetic diffusion region predicts, based 
on the high magnetic Lundquist number in the solar corona, 
subsonic horizontal inflows of plasma 
($v_{in} \approx 20$ km s$^{-1}$)
into the diffusion region and acceleration at the
X-point to Alfv\'enic vertical outflows 
($v_A \approx 2000$ km s$^{-1}$)
(in upward and downward directions), which  
yields a scaling law of $E_{diss} \propto B_{free}^2 L r^2$
(Eq.~22). The best-fit solution,
according to CCC=0.84 and $\alpha=1.08\pm0.05$ (Fig.~6c),
suggests that the flare volume scales with the 1D flare ribbon 
model $V \propto L r^2$, similar to the helical twisting model.
Thus, we learn that the actually dissipated magnetic energy
during solar flares is confined to chromospheric ribbons,
rather than to a coronal 3-D flare plasma volume, which
is the same dimensional topology as we found for the free energy.
This conclusion supports the thick-target model (with
chromospheric heating by precipitating electrons) and
contradicts direct heating in the reconnection region.} 

\item{The Rosner-Tucker-Vaiana (RTV) model is based on the
assumption that a constant volumetric heating rate is 
balanced by thermal conductive and radiative loss,
at the peak time of the flare, while heating is dominant
before the flare peak, and conductive and radiative losses
are dominant after the flare peak. This leads to an
equilibrium solution of $T_e \propto (p_0 L)^{1/3}$ for
a single loop with length $L_{th}$, pressure $p_0$, and
apex temperature $T_e$. The best match with data is 
found for the parameter of the thermal energy, with
$E_{th} \propto T_e^3 L_{th}^2$, according to a
cross-correlation coefficient of CCC=0.90 and a
regression slope of $\alpha = 0.95\pm0.04$ (Fig.~8e).
We learn which parameters ($L_{th}, E_{th}, n_e$) can
be used explicitly to confirm the RTV law,
while other physical parameters ($T_e, EM$) have too
small of a range (for this particular data set of
M and X-class flares) to corroborate the RTV scaling law.}

\item{The Shibata-Yokoyama scaling law is a combination
of the Petschek-type and RTV scaling laws, based on four
physical assumptions; (i) the flare temperature at the
loop apex is balanced by the heating rate and the
conductive cooling rate, (ii) the heating rate is
given by the reconnection rate in Petschek's theory,
(iii) the upper limit of the (thermal) gas pressure
is given by the magnetic energy density, and (iv) the
volume of the heated flare plasma is estimated from
a cubic geometry $V=L^3$. The best fit 
has CCC=0.87 and $\alpha=1.18\pm0.05$ (Fig.~9b).
Consistency with the data is obtained for the 
expression of the emission measure, 
$EM \propto L_{th}^{5/3} n_e^{2/3} T_e^{8/3}$ (Fig.~9b), and
the thermal length scale, 
$L_{th} \propto EM^{3/5} n_e^{-2/5} T_e^{-8/5}$ (Fig.~9d).
Similar to the RTV scaling law, some parameters 
($T_e, B_{np}$) have too small of a range to be useful
to test the Shibata-Yokoyama scaling law, while others
($EM, L_{th}$) confirm its match with data explicitly.}
\end{enumerate}  

These results corroborate the validity of flare
scaling laws for the five types of investigated
flare phenomena, in the sense that we found 
self-consistency with observational data. Our 
data set here is biased towards large flares 
(of GOES M- and X-class), for which the temperature
range ($T_e = 26\pm 6$ MK) as well as the
range of the peak magnetic field strength 
($B_{np}=2100 \pm 400$ G) is relatively
narrow, so that the scaling cannot be
retrieved. On the other side, scaling laws
expressed explicitly for energies, emission
measures, and length scales have suffiently
large parameter ranges (2-3 orders of magnitude) 
to constrain the power law indexes of scaling laws.

In summary, the self-consistency between observables
and theoretically predicted scaling laws reveals
the success or failure of well-posed physical 
flare models, while model-free fitting of power law 
indices in heuristic expressions $X = a^\alpha b^\beta ...$ 
rarely reveal an adequate physical model. For
instance, the best-fit result $F_{heat} \propto B^1 L^{-1}$
found in simulations of coronal heating by
Schrijver et al.~(2004) clashes with the result of 
$F_{heat} \propto B^2$ by D\'emoulin et al.~(2003), since
neither their physical units nor their power law exponents
can be reconciled. In other words, fitting a physical
model is preferable over evaluating model-free scaling
laws without an underlying physical model. For future
progress we recommend to increase narrow
parameter ranges (such as temperature, density, and 
magnetic field parameters) in order to optimize
flare scaling law tests.

\vskip1cm
{\sl Acknowledgements:}
We acknowledge software support by Samuel Freeland, Greg Slater,
and Mark Noga, and constructive comments from an anonymous reviewer.
Part of the work was supported by NASA contract NNG04EA00C of the
SDO/AIA instrument and the NASA STEREO mission under NRL contract
N00173-02-C-2035.


\section*{ References }

\def\ref#1{\par\noindent\hangindent1cm {#1}}

\ref{Altschuler, M.D. and Newkirk, G.Jr. 1969, SoPh 9, 131}
\ref{Aschwanden, M.J., Kosugi, T., Hudson, H.S., Wills, M.J.
	and Schwartz, R.A. 1996, ApJ 470, 1198} 
\ref{Aschwanden, M.J. 2007, AdSpR 39, 1867}
\ref{Aschwanden, M.J., Stern, R.A., and G\"udel, M. 2008, ApJ 672, 659}
\ref{Aschwanden, M.J. 2010, SoPh 262, 235}
\ref{Aschwanden, M.J. and Sandman, A.W. 2010, ApJSS 140:723}
\ref{Aschwanden, M.J. 2013, SoPh 287, 323}
\ref{Aschwanden, M.J. and Shimizu, T. 2013, ApJ 776, 132}
\ref{Aschwanden, M.J., Zhang, J., and Liu, K. 2013, ApJ 775:23}
\ref{Aschwanden, M.J., Xu, Y., and Jing, J. 2014a, ApJ 797:50 (Paper I)}
\ref{Aschwanden, M.J., Boerner, P., Ryan, D., Caspi, A., McTiernan, J.M.,
	and Warren, H.P. 2015a, ApJ 802:53 (Paper II)}
\ref{Aschwanden, M.J., Boerner, P., Caspi, A., McTiernan, J.M.,
	Ryan, D., and Warren, H. 2015b, SoPh 290, 2733}
\ref{Aschwanden, M.J. 2019a, ApJ 885:49 (Paper IX)}
\ref{Aschwanden, M.J. 2019c, ApJ 874:131}
\ref{Aschwanden, M.J. 2020, ApJ 895:134, (Paper X)}
\ref{Crosby, N.B., Aschwanden, M.J., and Dennis, B.R. 1993, SoPh 143, 275}
\ref{D\'emoulin, P., van Driel-Gesztelyi, L., Mandrini, C.H.,
	Klimchuk, J.A., and Harra, L. 2003, ApJ 586, 592} 
\ref{Garcia, H.A. 1998, ApJ 504, 1051}
\ref{Guo, Y., Pariat, E., Valori, G., Anfinogentov, S., Chen, F.,
	Georgoulis, M.K., Liu, Y., Moraitis, K., Thalmann, J.K.,
	and Yang, S. 2017, ApJ 840:40}
\ref{Hood, A.W. and Priest, E. 1979, SoPh 64, 303} 
\ref{Hood, A.W. and Priest, E. 1981, GApFD 17, 297}  
\ref{Kahler, S.W. 2013, ApJ 769:35}
\ref{Kliem, B., Linton, M., T\"or\"ok, T., and Karlicky, M. 2010,
	SoPh 266, 91}
\ref{Klimchuk, J.A. and Porter, L.J. 1995, Nature 377, 131}
\ref{Lemen, J.R., Title, A.M., Akin, D.J., et al. 2012, SoPh 275, 17}
\ref{Mandrini, C.H., Demoulin, P., and Klimchuk, J.A. 2000,
	ApJ 530, 999.}
\ref{Morales, L.F. and Charbonneau, P. 2008, GRL 35, 4108M}
\ref{Namekata, K., Sakaue, T., Watanabe, K., Asai, A., and Shibata, K.
	2017, PASJ 69, id.7}
\ref{Parker, E.N. 1963, ApJS, 8, 177}
\ref{Parker, E.N. 1988, ApJ 330, 474}
\ref{Pesnell, W.D., Thompson, B.J., and Chamberlin, P.C.  2012, SoPh 275, 3}
\ref{Petschek, H.E. 1964, in Proc. AAS-NASA Symp., The Physics of Solar
	Flares, ed. W.N. Hess (Washington, DC: NASA), 425}
\ref{Press, W.H., Flannery, B.P., Teukolsky, S.A., and
	Vetterlikng, W.T. 1986, {\sl Numerical Recipes,
        The Art of Scientific Computing}, Cambridge University Press,
        Cambridge, p.304}
\ref{Pugh, C.E., Broomhall, A.M., and Nakariakov, V.M. 2019, A\&A 624, A65}
\ref{Priest, E.R. 1975, SoPh 43, 177}
\ref{Rosner, R., Tucker, W.H., and Vaiana, G.S. 1978, ApJ 220, 643} 
\ref{Sakurai, J.I. 1982, SoPh 76, 301}
\ref{Scherrer, P.H., Schou, J., Bush, R.J. 2012, SoPh 275, 207}
\ref{Schmidt, H.U. 1964, NASA SP 50, (ed. W.N. Hess), Washington DC} 
\ref{Schrijver, C.J., Sandman, A.W., Aschwanden, M.J.,
	and DeRosa, M.L. 2004, ApJ 615, 512} 
\ref{Shibata, K. and Yokoyama, T. 1999, ApJ 526, L49}
\ref{Shibata, K. and Yokoyama, T. 2002, ApJ 577, 432}
\ref{Sweet, P.A. 1958, in IAU Symp. 6, Electromagnetic Phenomena
	in Cosmic Physics, (ed. B. Lehnert (Cambridge: Cambridge
	Univ. Press), 123}
\ref{T\"or\"ok, T. and Kliem, B. 2003, A\&A 406, 1043}
\ref{Vekstein, G. and Katsukawa, Y. 2000, ApJ 541, 1096}
\ref{Warmuth, A. and Mann, G. 2016a, A\&A 588, A115}
\ref{Warmuth, A. and Mann, G. 2016b, A\&A 588, A116}
\ref{Wiegelmann, T., Inhester, B., and Sakurai, T. 2006, SoPh 223, 215}
\ref{Yamamoto, T.T., Shiota, D., Sakajiri, T., Akiyama, S., Isobe, H.,
	and Shibata, K. 2002, ApJ 579, 45}
\ref{Yokoyama, T. and Shibata, K. 1998, ApJ 494, 113} 
 
\clearpage


\begin{table}[t]
\tabletypesize{\normalsize}
\caption{Observational parameters of $N=173$ M- and X-class flare events analyzed
in Aschwanden, Xu, and Jing (2014).}
\medskip
\begin{tabular}{lrrrr}
\hline
                                                        & Minimum  & Maximum    & Mean          & Median \\
\hline
\hline
Nonpotential magnetic energy $E_{np} [10^{30}$ erg]
 &     86.0 &     4866.0 &   1508.6$\pm$  1063.0 &   1146.0\\
Potential magnetic energy $E_{p} [10^{30}$ erg]
 &     85.0 &     3949.0 &   1390.1$\pm$   946.4 &   1135.0\\
Free energy $E_{free} [10^{30}$ erg]
 &      1.0 &      951.0 &    118.5$\pm$   160.8 &     64.0\\
Dissipated magnetic energy $E_{diss} [10^{30}$ erg]
 &      1.0 &     1546.0 &    181.2$\pm$   249.2 &    110.0\\
Nonpotential field strength $B_{np}$ [G]
 &    344.0 &     3403.0 &   2036.3$\pm$   366.7 &   2075.9\\
Potential field strength $B_p$ [G]
 &    329.0 &     3214.0 &   1964.7$\pm$   336.7 &   1992.0\\
Free energy field strength $B_{free}$ [G]
 &    100.5 &     1227.8 &    497.3$\pm$   246.0 &    453.7\\
Dissipated energy field strength $B_{diss}$ [G]
 &    111.7 &     1868.8 &    637.2$\pm$   300.4 &    566.0\\
Length scale  $L_{mag}$ [Mm]
 &     13.0 &      236.0 &     40.2$\pm$    25.0 &     33.0\\
Sunspot radius  r[Mm]
 &      1.4 &        9.1 &      4.5$\pm$     1.1 &      4.3\\
Helical twisting radius  R[Mm]
 &      1.7 &       74.9 &     11.9$\pm$    11.9 &      7.9\\
Depth of magnetic charges  d[Mm]
 &      1.8 &       11.8 &      5.8$\pm$     1.4 &      5.6\\
Flare duration  $\tau_{flare}$ [s]
 &    288.0 &    14760.0 &   1498.1$\pm$  1812.2 &    900.0\\
Thermal length scale $L_{th}$ [Mm]
 &     10.8 &      282.1 &     82.0$\pm$    48.4 &     68.5\\
Emission measure-weighted electron temperature $T_e$ [MK]
 &      6.2 &       41.6 &     26.1$\pm$     5.6 &     26.4\\
Mean flare electron density  $n_e [10^{10}$ cm$^{-3}$]
 &      2.8 &       56.2 &     10.3$\pm$     7.4 &      8.5\\
Total thermal emission measure EM [$10^{48}$ cm$^3$]
 &      0.4 &      182.0 &     17.0$\pm$    23.6 &      9.5\\
Total thermal energy $E_{th}$ [$10^{30}$ erg]
 &      0.2 &      215.3 &     16.6$\pm$    29.8 &      7.0\\
\hline
\end{tabular}
\end{table}


\begin{table}
\caption{Scaling relationships are listed for 18 different
scaling law models, specified by the model name, the
figure number, the dimension of the length scale,
the scaling law, the normalization factor $q_{norm}$,
the cross-correlation coefficient CCC, and the linear 
regression slope $\alpha$.}
\begin{tabular}{lllllll}
\hline
Model   & Figure & Dimension     & Scaling                       & Norm   & Coefficient & Slope     \\
        &	 & $L$, $L_{th}$ & Relationship                  & factor & CCC         & $\alpha$  \\
\hline
\hline
EP-3D   & 4a     & 3    & $E_p \propto B_p^2 L^3$           & 0.2    & 0.70      & 1.37$\pm$0.11\\
EP-2D   & 4b     & 2    & $E_p \propto B_p^2 L^2 r$         & 6.3    & 0.69      & 1.11$\pm$0.09\\
EP-1D   & 4c     & 1    & $E_p \propto B_p^2 L r^2$         & 178    & 0.62      & 0.85$\pm$0.08\\
\hline
EF-3D   & 5a     & 1    & $E_{free} \propto B_p^2 R^2 L$    & 0.23   & 0.86      & 1.33$\pm$0.06\\
EF-2D   & 5b     & 0    & $E_{free} \propto B_p^2 R^2 r$    & 6.0    & 0.88      & 1.19$\pm$0.05\\
EF-1D   & 5c     & -1   & $E_{free} \propto B_p^2 R^2 L^{-1} r^2$ & 165 & 0.86 & 1.04$\pm$0.05\\
\hline
ED-3D   & 6a     & 3    & $E_{diss} \propto B_{free}^2 L^3$   & 0.23   & 0.88      & 1.50$\pm$0.06\\
ED-2D   & 6b     & 2    & $E_{diss} \propto B_{free}^2 L^2 r$ & 7.61   & 0.88      & 1.29$\pm$0.05\\
ED-1D   & 6c     & 1    & $E_{diss} \propto B_{free}^2 L r^2$ &  242   & 0.84      & 1.08$\pm$0.05\\
\hline
RTV-T   & 8a	 & 1/2  & $T_e \propto n_e^{1/2} L_{th}^{1/2}$  & 1.24   & 0.34    & 0.18$\pm$0.04\\
RTV-L   & 8b	 & 1    & $L_{th} \propto T_e^{3} n_e^{-1}$ & 0.68   & 0.68        & 0.78$\pm$0.06\\
RTV-N   & 8c     & -1 	& $n_e \propto T_e^{2} L_{th}^{-1}$ & 0.65   & 0.80        & 1.15$\pm$0.07\\
RTV-EM	& 8d     & 1    & $EM  \propto T_e^4 L_{th}$	    & 0.0016 & 0.56        & 0.69$\pm$0.08\\   
RTV-ETH & 8e     & 2    & $E_{th} \propto T_e^3 L_{th}^2$   & 0.015  & 0.90        & 0.95$\pm$0.04\\
\hline
SY-T	& 9a     & 2/7  & $T_e \propto B^{6/7} n_e^{-1/7} L_{th}^{2/7}$   & 0.12 & -0.02  & -0.02$\pm$0.09\\
SY-EM   & 9b 	 & 5/3  & $EM \propto L_{th}^{5/3} n_e^{2/3} T_e^{8/3}$   & 0.0014 & 0.76 & 0.79$\pm$0.05\\
SY-B    & 9c 	 & ...  & $B_{np} \propto EM^{-1/5} n_e^{3/10} T_e^{17/10}$&3.32 & 0.08   & 0.19$\pm$0.18\\
SY-L    & 9d  	 & 1    & $L_{th} \propto EM^{3/5} n_e^{-2/5} T_e^{-8/5}$ &   48 & 0.87   & 1.18$\pm$0.05\\
\hline
\hline
\end{tabular}
\end{table}


\begin{figure}
\centerline{\includegraphics[width=1.0\textwidth]{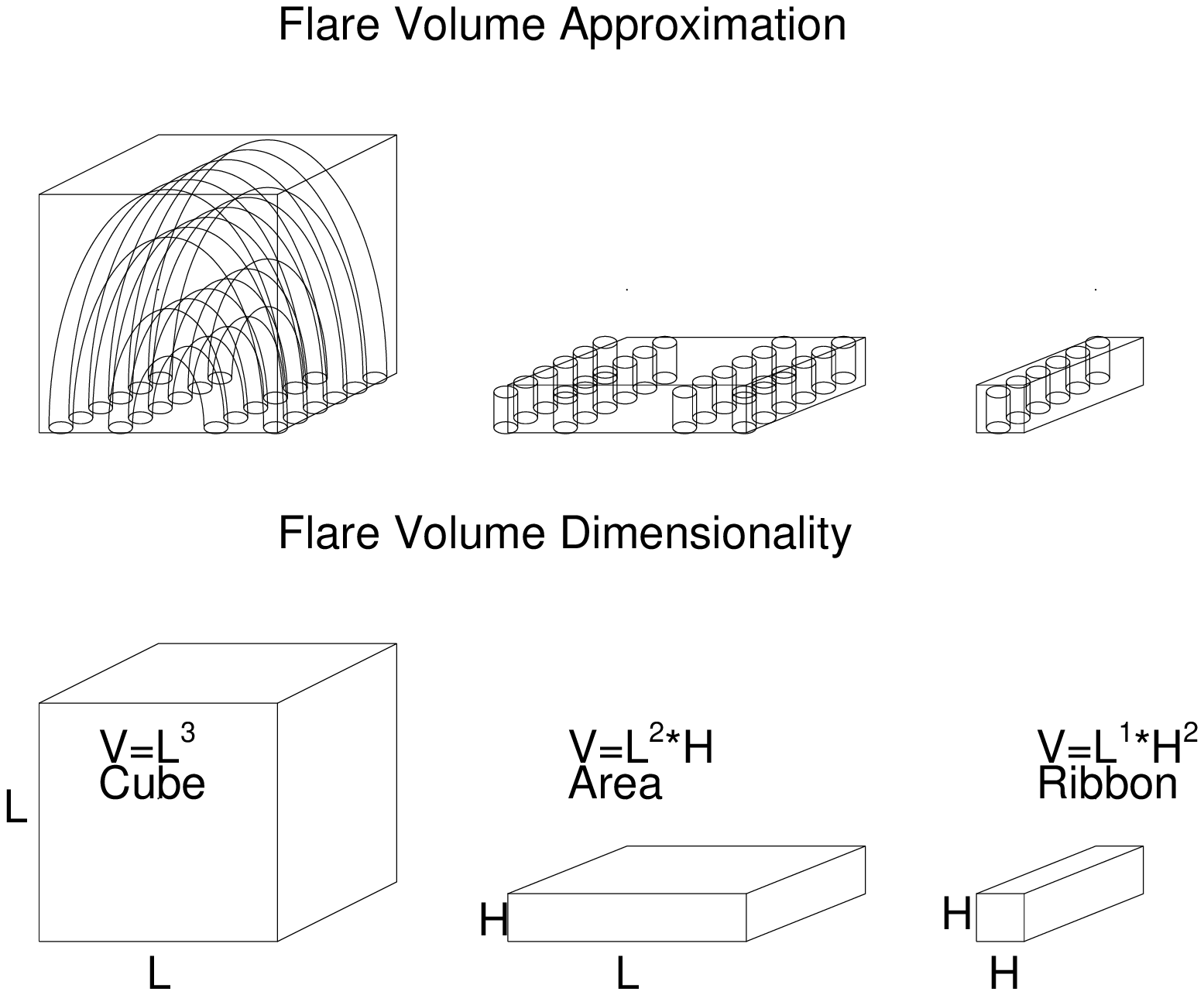}}
\caption{The dimensionality of the flare volume in scaling
laws is depicted, which may be approximated with a 3D cube
encompassing a flare loop arcade (left), a 2D flare area with
a small height extent $H$ corresponding to the magnetic scale 
height (middle), or a 1D ribbon with cross-section $H^2$. 
Note that each of these candidate geometric models has a 
different dependence of the flare volume on the length 
variable $L$, i.e., $V(L) \propto L^3, L^2$, or $L$.}
\end{figure}

\begin{figure}
\centerline{\includegraphics[width=1.0\textwidth]{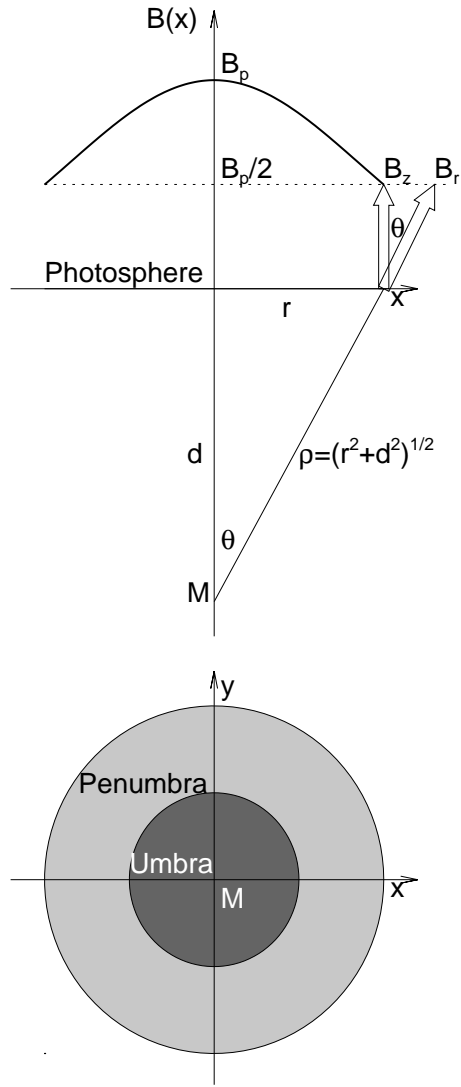}}
\caption{Schematic of sunspot parameters from a top view in the 
$[x,y]$ plane, showing the umbra and penumbra (bottom panel),
a side view with the location of the magnetic charge $M$
at sunspot radius $r$ and subphotospheric depth $d$ (middle
panel), and the potential field strength $B(x)$
as a function or the radius $x$, with the sunspot radius $r$
defined by the value of the half maximum peak $B_p/2$ (top
panel). $B_r$ is the radial field component, and $B_z$ is
the line-of-sight component measured by the observer.} 
\end{figure}

\begin{figure}
\centerline{\includegraphics[width=0.8\textwidth]{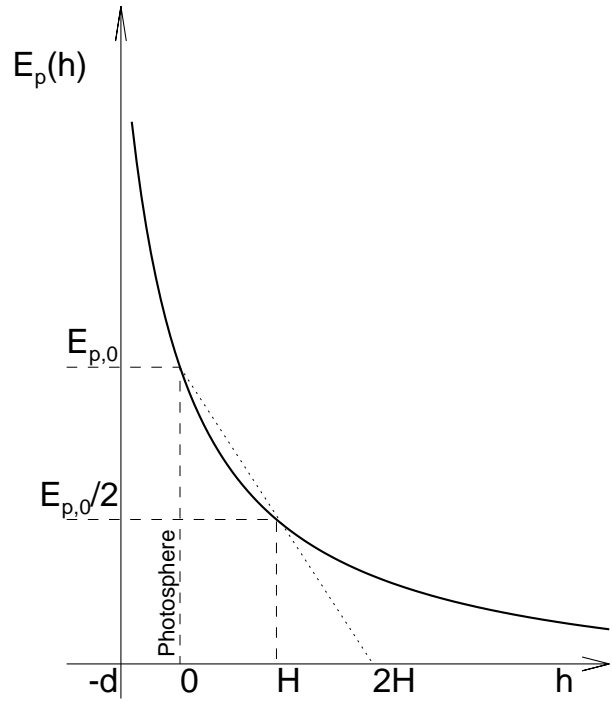}}
\caption{Schematic of height dependence of the magnetic field
energy $E_p(h)$, with the magnetic charge located at a 
sub-photospheric depth $h=-d$, the photospheric
height at $h=0$ with the peak energy value $E_{p,0}$, 
and the half peak value $E_{p,0}/2$ at $h=H$.}
\end{figure}

\begin{figure}
\centerline{\includegraphics[width=1.0\textwidth]{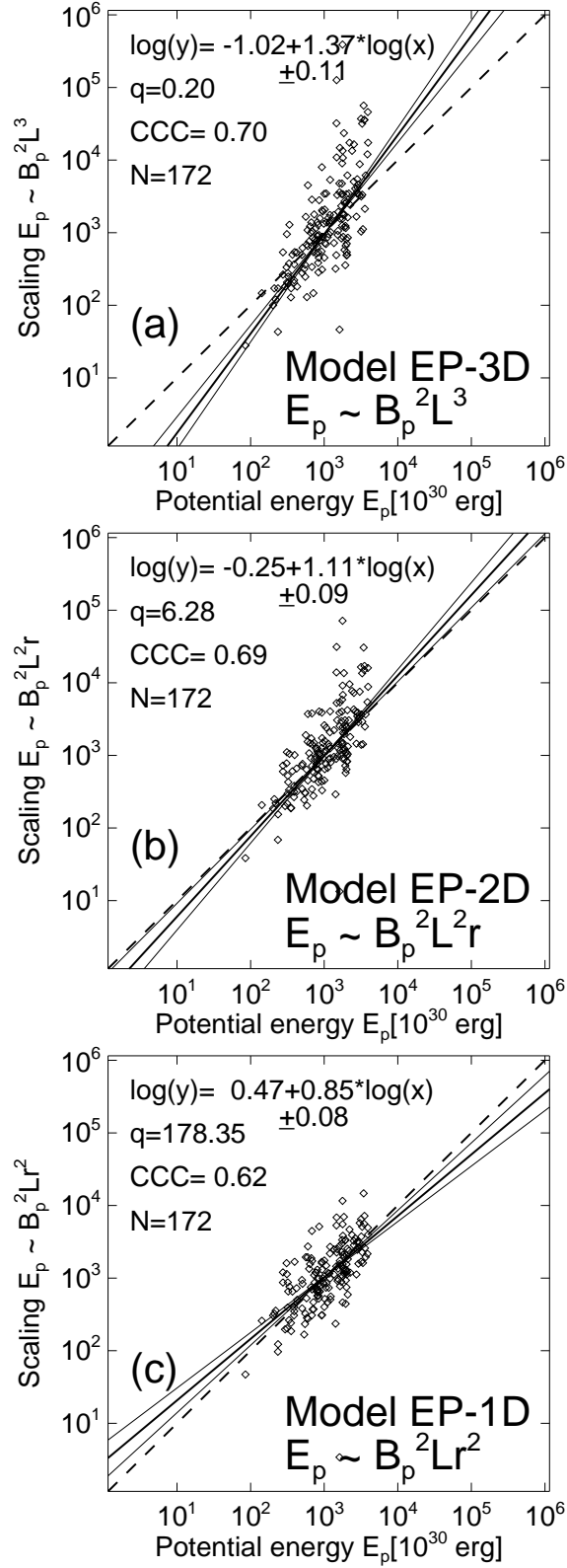}}
\caption {Parameter correlations are shown for theoretrical
scaling relationships $E_p^{theo}$ (Y-axis) of the potential
field energy of a flaring active region as a function of the 
observed values $E_p^{obs}$, for $N=173$ M- and X-class flare
events (diamond symbols), along with linear regression fits,
the best fit $Y(X)$ (thick solid line), standard deviations 
(thin solid lines), and equivalence (dashed lines). 
The three cases correspond to models with different 
flare geometries with dimensions of $V \propto L^3$ (a), 
$V \propto L^2$ (b), and $V \propto L$ (c), as depicted
in Fig.~1 and listed as models EP-3D, EP-2D, and EP-1D
in Table 2.} 
\end{figure}

\begin{figure}
\centerline{\includegraphics[width=1.0\textwidth]{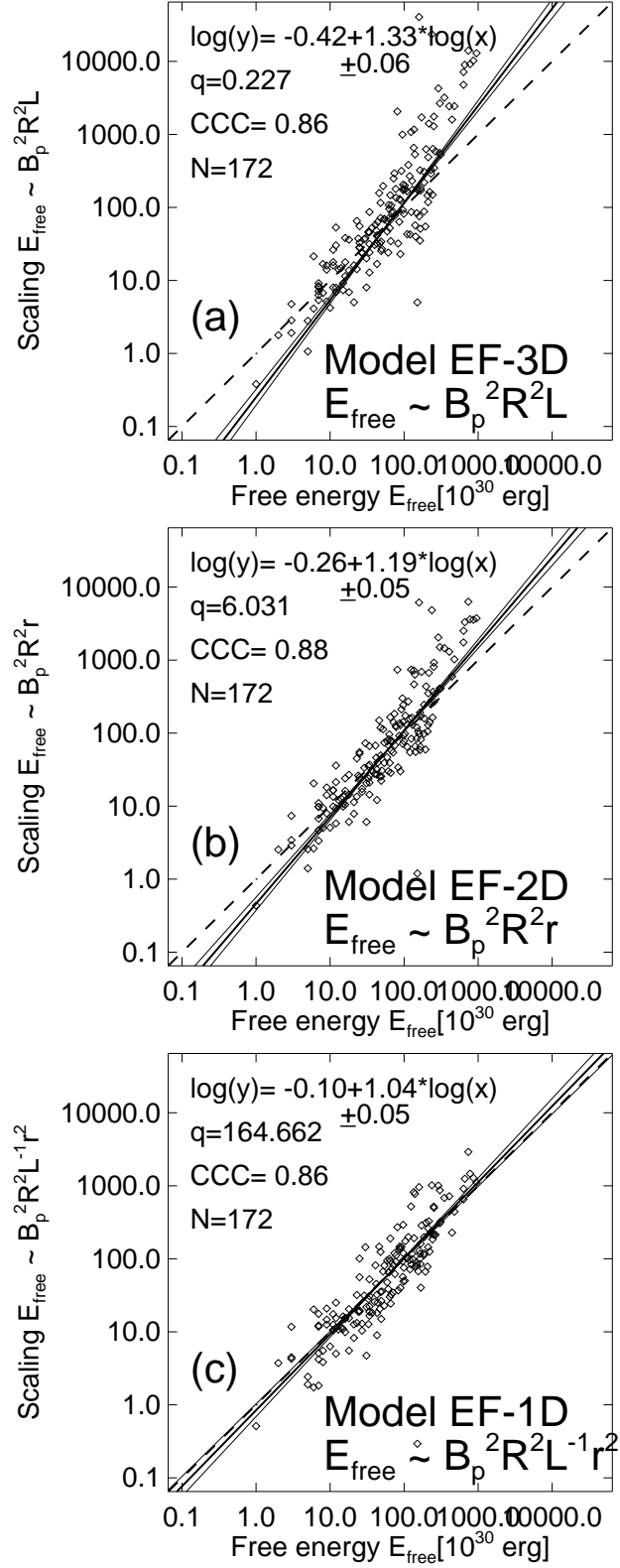}}
\caption {Parameter correlations are shown for the theoretrical
scaling relationships $E_{free}^{theo}$ (Y-axis) of the free magnetic
energy as a function of the observed values $E_{free}^{obs}$ for 
$N=173$ M- and X-class flare events (diamond symbols). 
Otherwise similar representation as in Fig.~4.}
\end{figure}

\begin{figure}
\centerline{\includegraphics[width=1.0\textwidth]{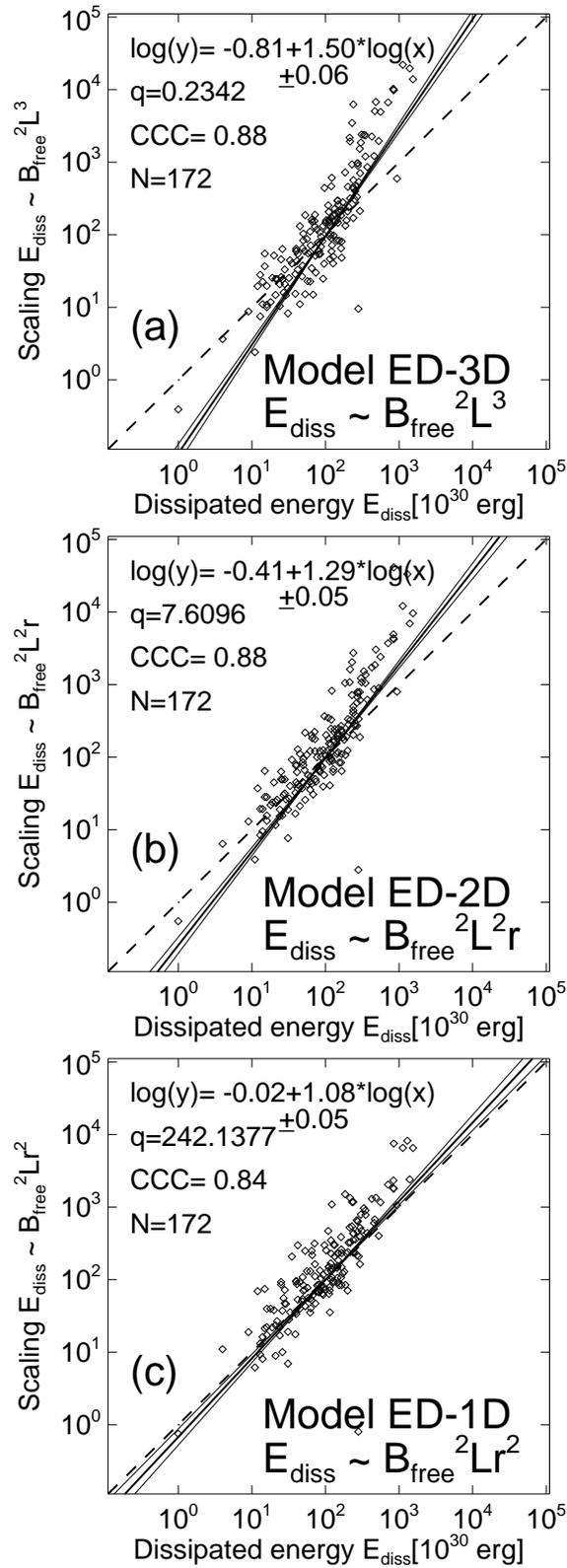}}
\caption {Parameter correlations of 3 Petschek-type models
are shown for the theoretrical scaling relationships 
$E_d^{theo}$ (Y-axis) of the dissipated magnetic
energy as a function of the observed values $E_d^{obs}$ for 
$N=173$ M- and X-class flare events (diamond symbols). 
Otherwise similar representation as in Fig.~5.}
\end{figure}

\begin{figure}
\centerline{\includegraphics[width=1.0\textwidth]{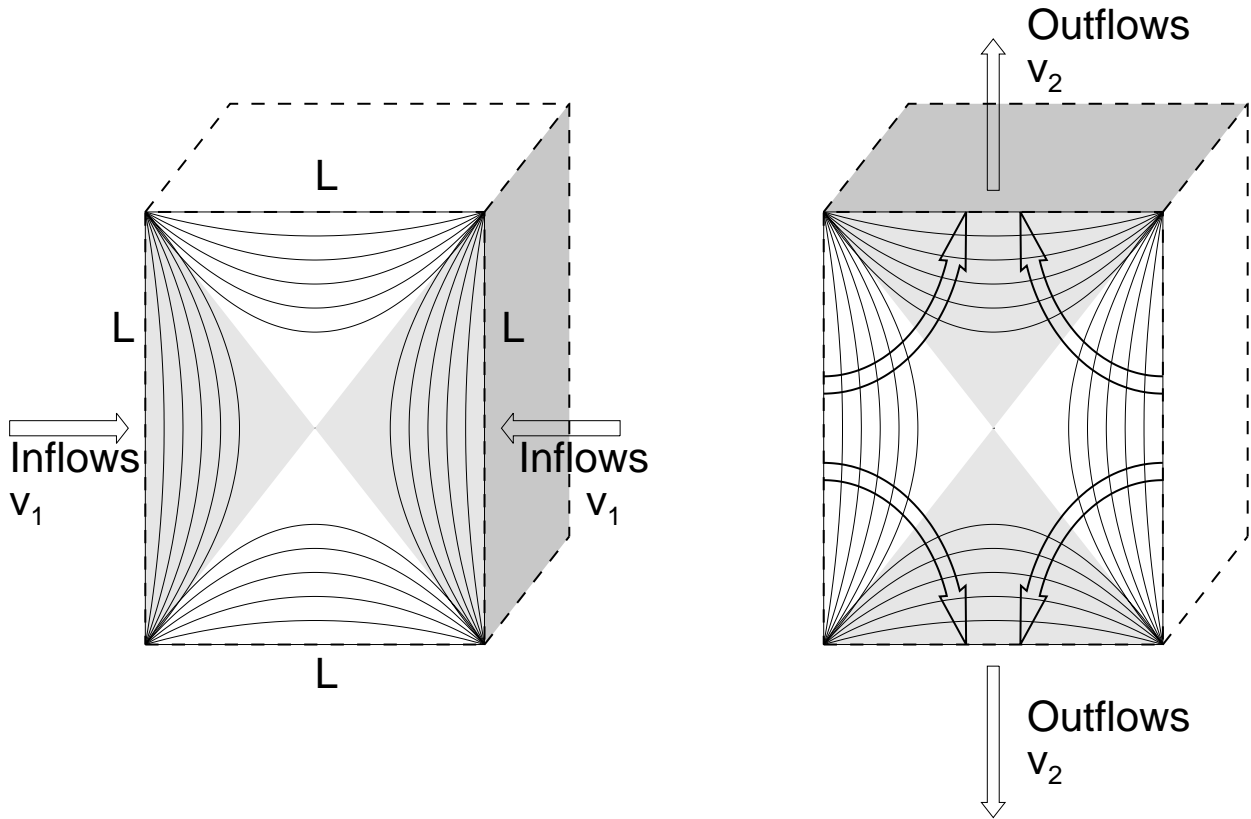}}
\caption{A schematic of the Petschek magnetic reconnection
process, showing the horizontally directed inflows $v_1$
that drive the reconnection (left panel), and the vertical 
outflows $v_2$ in upward and downward direction from the central
X-point (right panel). The 3D box (dashed lines) with a
length scale $L$ in all directions marks the magnetic 
diffusion region.}
\end{figure} 

\begin{figure}
\centerline{\includegraphics[width=1.0\textwidth]{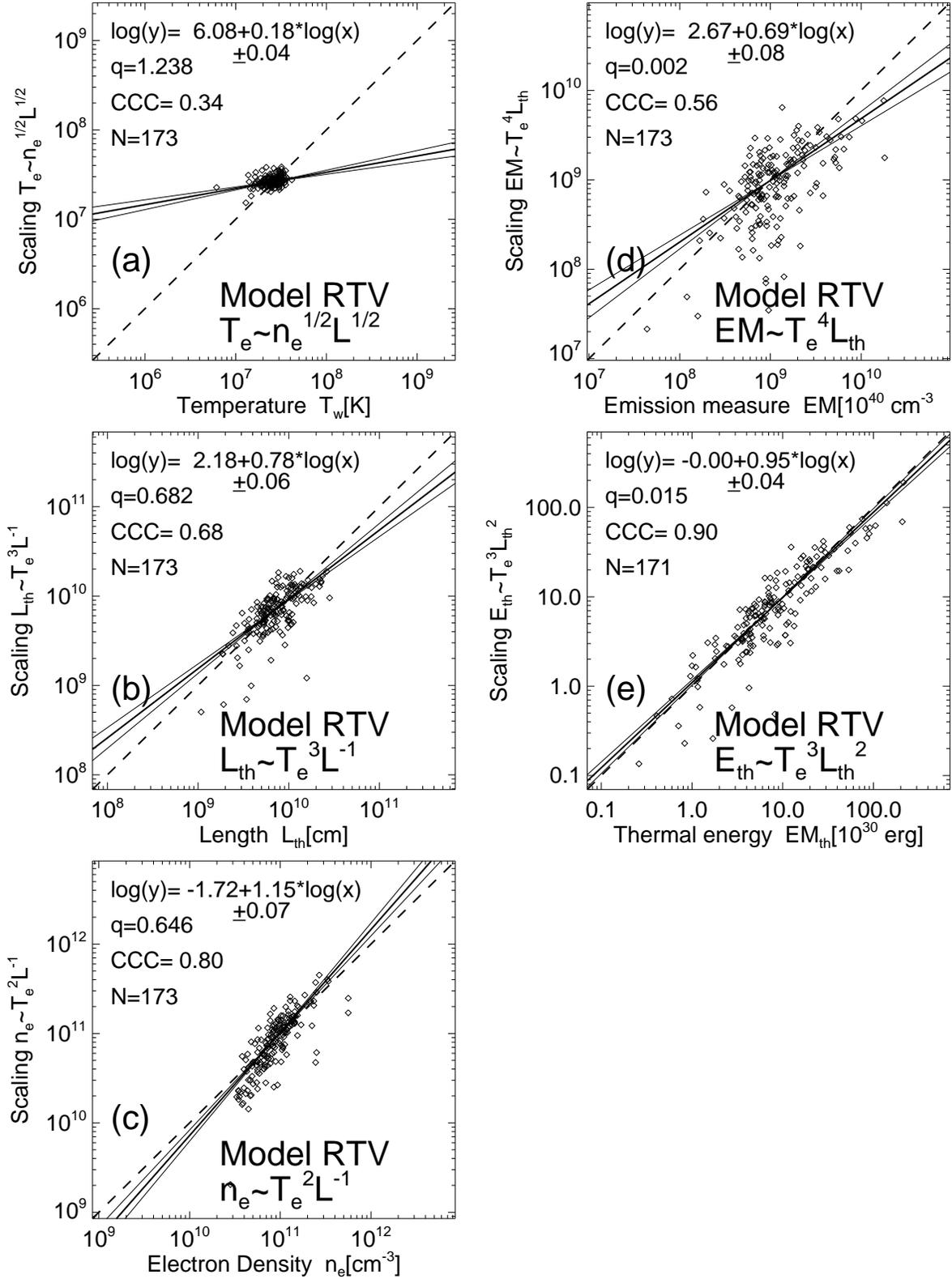}}
\caption {Correlations of five Rosner-Tucker-Vaiana
(RTV) parameters are shown in form of the observed values
on the X-axis versus the theoretrically predicted scaling 
relationships on the Y-axis, for N=193 M- and X-class 
flare events (diamond symbols). 
Otherwise similar representation as in Figs.~(4)-(6).}
\end{figure}

\begin{figure}
\centerline{\includegraphics[width=1.0\textwidth]{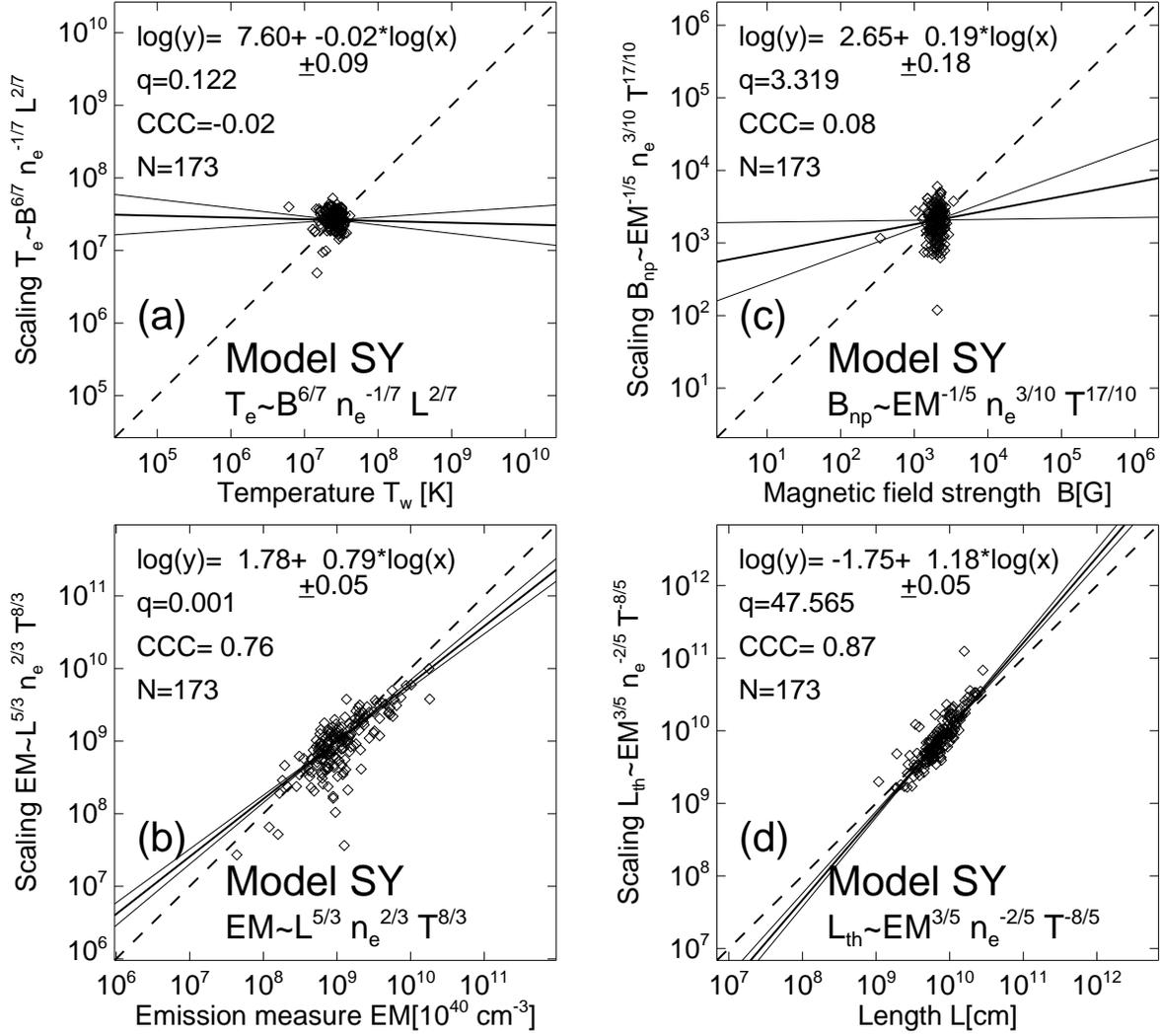}}
\caption {Correlations of four parameters of the
Shibata-Yokoyama reconnection model are shown 
in form of the observed values
on the X-axis versus the theoretrically predicted scaling 
relationships on the Y-axis, for $N=173$ M- and X-class 
flare events (diamond symbols). 
Otherwise similar representation as in Figs.~(4)-(6).}
\end{figure}

\end{document}